# A Large Open Access Dataset of Brain Metastasis 3D Segmentations with Clinical and Imaging Feature Information

Divya Ramakrishnan[1], Leon Jekel[1,2], Saahil Chadha[1], Anastasia Janas[1,3], Harrison Moy[1,4], Nazanin Maleki[1], Matthew Sala[1,5], Manpreet Kaur[1,6], Gabriel Cassinelli Petersen[1,7], Sara Merkaj[1,8], Marc von Reppert[1,9], Ujjwal Baid[10,11], Spyridon Bakas[10,11], Claudia Kirsch[1,12,13], Melissa Davis[1], Khaled Bousabarah[14], Wolfgang Holler[14], MingDe Lin[1,15], Malte Westerhoff[14], Sanjay Aneja[16,17], Fatima Memon[1], and Mariam S. Aboian[1]

**Affiliations**

1. Yale School of Medicine, Department of Radiology and Biomedical Imaging, New Haven, CT, USA
2. University of Essen School of Medicine, Essen, Germany
3. Charité University School of Medicine, Berlin, Germany
4. Wesleyan University, Middletown, CT, USA
5. Tulane University School of Medicine, New Orleans, LA, USA
6. Ludwig Maximilian University School of Medicine, Munich, Germany
7. University of Göttingen School of Medicine, Göttingen, Germany
8. Ulm University School of Medicine, Ulm, Germany
9. University of Leipzig School of Medicine, Leipzig, Germany
10. Division of Computational Pathology, Department of Pathology & Laboratory Medicine, Indiana University School of Medicine, Indianapolis, IN, USA
11. Department of Radiology and Department of Pathology & Laboratory Medicine, Perelman School of Medicine, University of Pennsylvania, Philadelphia, PA, USA
12. School of Clinical Dentistry, University of Sheffield, Sheffield, England
13. Diagnostic, Molecular and Interventional Radiology, Biomedical Engineering Imaging, Mount Sinai Hospital, New York City, NY, USA
14. Visage Imaging, GmbH, Berlin, Germany
15. Visage Imaging, Inc., San Diego, CA, USA
16. Department of Therapeutic Radiology, Yale School of Medicine, New Haven, CT, USA
17. Center for Outcomes Research and Evaluation (CORE), Yale School of Medicine, New Haven, CT, USA

Corresponding author: Mariam Aboian (mariam.aboian@yale.edu)

## Abstract

Resection and whole brain radiotherapy (WBRT) are the standards of care for the treatment of patients with brain metastases (BM) but are often associated with cognitive side effects. Stereotactic radiosurgery (SRS) involves a more targeted treatment approach and has been shown to avoid the side effects associated with WBRT. However, SRS requires precise identification and delineation of BM. While many AI algorithms have been developed for this purpose, their clinical adoption has been limited due to poor model performance in the clinical setting. Major reasons for non-generalizable algorithms are the limitations in the datasets used for training the AI network. The purpose of this study was to create a large, heterogenous, annotated BM dataset for training and validation of AI models to improve generalizability. We present a BM dataset of 200 patients with pretreatment T1, T1 post-contrast, T2, and FLAIR MR images. The dataset includes contrast-enhancing and necrotic 3D segmentations on T1 post-contrast and whole tumor (including peritumoral edema) 3D segmentations on FLAIR. Our dataset contains 975 contrast-enhancing lesions, many of which are sub centimeter, along with clinical and imaging feature information. We used a streamlined approach to database-building leveraging a PACS-integrated segmentation workflow.

## Background & Summary

Brain metastases (BM) develop in up to 30-40% of patients with a primary malignancy, particularly those with lung cancer, breast cancer, and melanoma.[1,2] Palliative treatment for BM includes resection, whole brain radiotherapy (WBRT), and, more recently, stereotactic radiosurgery (SRS).[1] Although WBRT can reduce the neurological symptoms of BM, the overall survival has been shown to be decreased in patients with certain risk factors, including older age, lower baseline cognitive performance status, and >3 BM.[3,4] SRS provides a more targeted and less toxic approach to BM treatment than WBRT and can be performed when patients present with >10 lesions although its predominant use is still in treatment of localized metastatic disease.[5,6] In fact, one meta-analysis revealed a significant improvement in performance status and local control in patients treated with WBRT plus SRS compared to WBRT alone.[7] Localization and accurate delineation of BM margins are critical for effective SRS treatment.[8] In addition, differentiation of BM from high-grade gliomas, such as glioblastoma, can be challenging, and textual analysis of the peritumoral environment on T2/FLAIR sequences can aid in differentiation of these tumor subtypes.[9]

To address the challenge of BM diagnosis and delineation, several artificial intelligence (AI) tools, including machine learning (ML) and deep learning (DL) algorithms, have been developed in the past decade.[8,10–14] While many of these algorithms showed promising results in BM diagnosis and auto-segmentation, there is still a large gap in the clinical implementation and adoption of these algorithms.[12,15] One reason for this gap is the lack of algorithm generalizability to real-world datasets. In fact, many algorithms are trained and developed on small single-institution hospital datasets that lack diversity in patient populations and imaging protocols, which are often present in the clinical setting.[12] In fact, one meta-analysis of BM algorithms revealed that the average sample size of datasets used to train algorithms was around 150, with half of the studies explicitly including patients with only solitary BM.[12] Thus, there is a critical need for large, diverse, and open-access datasets to better train AI algorithms and to challenge AI models to perform accurate assessments on a large breadth of patient cases.[12] To date, there are only two publicly available BM datasets, both of which contain under 200 patients with pretreatment segmentations solely on T1 post-contrast.[16,17]

We curated a dataset of 200 patients with a clinical or pathological diagnosis of BM with accompanying clinical and qualitative/quantitative imaging data. In addition to enhancing tumor 3D segmentations, our dataset also provides 3D segmentations of necrotic tumor portions on T1 post-contrast and whole tumor (including peritumoral edema) on FLAIR. Our dataset includes several sub-centimeter contrast-enhancing lesions, which are critical for training algorithms to recognize subtle lesions on imaging. Manual 3D tumor segmentations using a commercially available semi-automatic segmentation tool was performed in a novel workflow directly in a research instance of our PACS (AI Accelerator, Visage Imaging, Inc., San Diego, CA),[18] which allowed for the creation and validation of segmentations in an accelerated time frame. We are in the process of making our dataset publicly available with all tumor segmentations (contrast-enhancing, necrotic, and peritumoral edema), standard MRI sequences (T1, T1 post-contrast, T2, and FLAIR), and an Excel file containing clinical information and qualitative/quantitative imaging features. We hope that our dataset contributes to the training and validation of future BM AI algorithms with the goal of their implementation, translation, and adoption in clinical practice for BM diagnosis and treatment.

## Methods

*Subject characteristics.* Patients were queried from the Yale New Haven Hospital (YNHH) database from 2013 to 2021, the YNHH tumor board registry in 2021, and the YNHH Gamma Knife registry from 2017 to 2021. Inclusion criteria were a clinical or pathological diagnosis of brain metastasis confirmed on the electronic medical record and availability of all four pretreatment standard MRI sequences (T1, T1 post-contrast, T2, and FLAIR) without significant motion artifact. There was a total of 200 patients included in the dataset. Of the 200 patients, the following was the breakdown of primary tumor origin: non-small cell lung cancer (86, 43%), melanoma (41, 20.5%), breast cancer (26, 13%), small cell lung cancer (17, 8.5%), renal cell carcinoma (16, 8%), and gastrointestinal cancers (14, 7%).

***Image acquisition.*** A summary of all imaging parameters for FLAIR and T1 post-contrast images of the 200 patients can be found in Table 1. The images were obtained on 1-T (4, 2%), 1.5-T (113, 56.5%), and 3-T (83, 41.5%) MRI scanners. Scanner vendors included Siemens (158, 79%), General Electric (31, 15.5%), Philips (7, 3.5%), and Hitachi (4, 2%).

| Imaging Parameter | FLAIR | T1 post-contrast |
| --- | --- | --- |
| Acquisition (n, %) | 2D (193, 96.5%)<br>3D (5, 2.5%)<br>N/A (2, 1%) | 2D (32, 16%)<br>3D (166, 83%)<br>N/A (2, 1%) |
| Median (range) echo time (msec) | 92.0 (10.0 - 400.0) | 3.1 (1.8 - 26.1) |
| Median (range) repetition time (msec) | 9000.0 (1700.0 - 12000.0) | 1900.0 (5.9 - 2619.8) |
| Median (range) slice thickness (mm) | 5.0 (1.0 - 5.5) | 1.0 (0.9 - 5.0) |
| Median (range) slice spacing (mm) | 5.0 (0.0 - 7.5) | 0.0 (0.0 - 7.0) |

**Table 1.** Summary of imaging parameters for FLAIR and T1 post-contrast sequences. *N/A = not available; range = minimum – maximum

***Segmentation procedure.*** The DICOM studies for all 200 patients were sent and de-identified from the clinical production (Visage 7, Visage Imaging, Inc., San Diego, CA) to a research instance of our PACS. To streamline the segmentation workflow, a custom hanging protocol and eight-viewer layout were designed to automatically 3D register and display the relevant MR imaging sequences upon study load.[18,19] Manual segmentations were performed by one medical student (L.J.) on the research PACS using a commercially available semi-automatic 3D segmentation tool (Fig 1).[18]

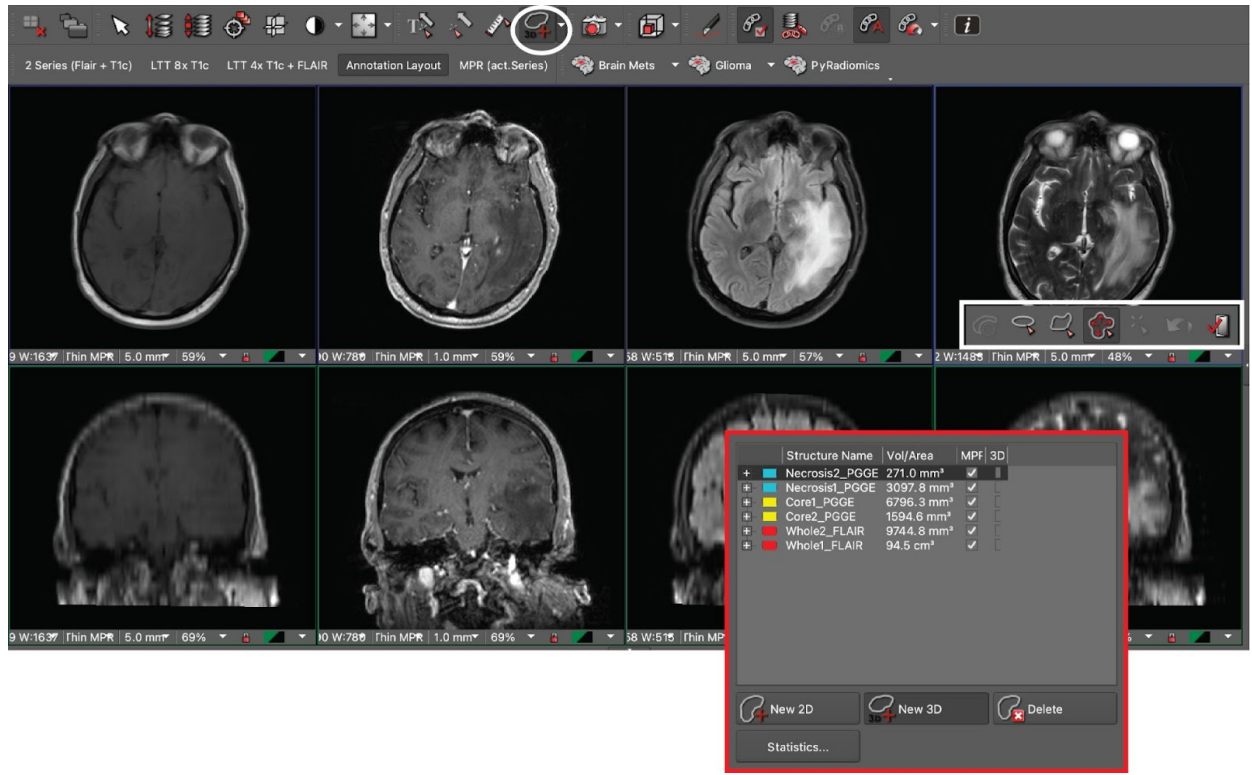

**Fig 1.** Research PACS annotation layout with T1, T1 post-contrast, FLAIR, and T2 sequences aligned. The PACS interface incorporates a 3D volumetric tool (white circle / rectangle) and displays labeled segmentations for two metastases in the display window (red rectangle).

The segmentations were checked and manually revised as needed by two board-certified neuroradiologists (M.S.A. and F.M.) with more than seven years of clinical experience each. Contrast-enhancing lesions and necrotic portions were segmented on T1 post-contrast (Fig 2A). In total, there were 975 contrast-enhancing lesions among all patients with 285 patients having necrotic components. Whole tumor (including peritumoral edema) was segmented on FLAIR

(Fig 2B). A total of 662 lesions had peritumoral edema surrounding contrast-enhancement. Notably, because a 3D registration of the various MR imaging sequences was performed using the custom hanging protocol, the segmentation masks could be accurately copied and pasted between MR imaging sequences.[18]

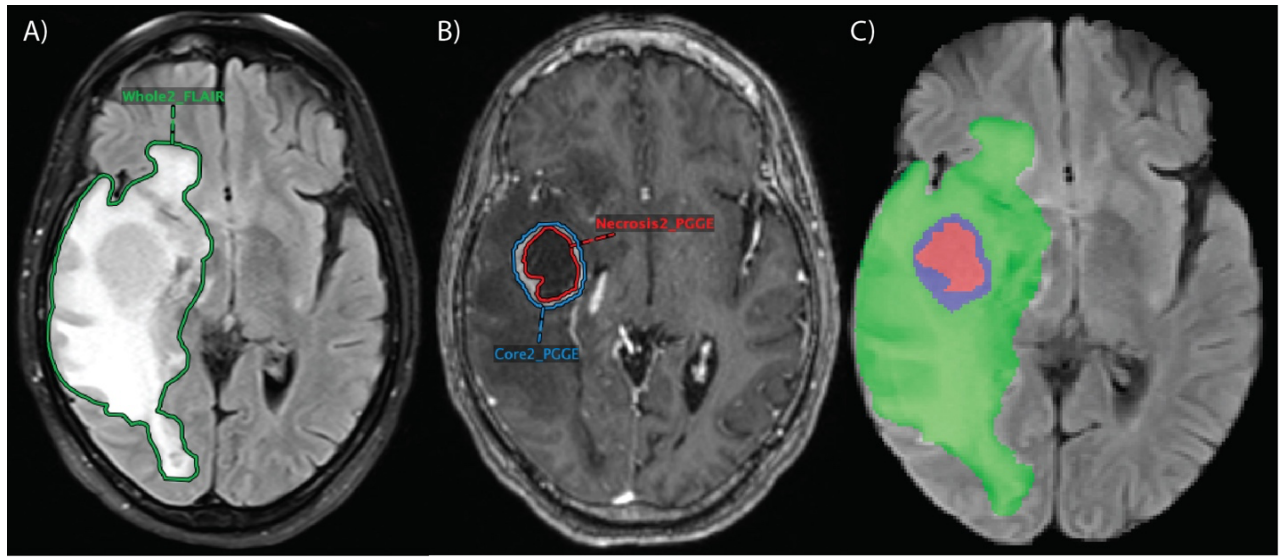

**Fig 2.** Segmentations performed in research PACS for one patient. FLAIR (A) included segmentation of whole tumor ("Whole2_FLAIR"), including peritumoral edema, and T1 post-contrast (B) included segmentations of the contrast-enhancing lesion ("Core2_PGGE") and necrotic portions within the contrast-enhancing lesion ("Necrosis2_PGGE"). (C) Combined segmentation mask overlaid on FLAIR sequence in NIfTI format with the green region representing peritumoral edema, the blue region representing contrast-enhancing tumor, and the red region representing necrotic tumor.

*Clinical data and anonymization.* Clinical data for all patients were collected from the electronic medical record. They include the following: age at diagnosis, sex, ethnicity, smoking history in pack-years, primary tumor origin, presence of extranodal metastasis, and time to death or last note in the electronic medical record as of July 2022. The following qualitative/quantitative imaging features were included: presence of infratentorial involvement, total number of lesions (contrast-enhancing, necrotic, and peritumoral edema), total volume of all regions (contrast-enhancing, necrotic, and peritumoral edema), ratio of necrotic to contrast-enhancing volume, and ratio of peritumoral edema to contrast-enhancing volume.

De-identification was implemented on the research server and occurred directly upon receipt of the DICOM images from either the PACS production system or the long-term archive. No non-anonymized images were stored on the research server. The de-identification removes/modifies all metadata that have identifiable information according to the DICOM standard PS3.15 2018b Appendix E "Attribute Confidentiality Profiles". Specifically, the "Basic Profile" combined with the "Clean Descriptors Option", the "Clean Structured Content Option" and the "Retain Longitudinal Temporal Information with Modified Dates Option" were implemented. The PatientID, Accession number, and StudyInstanceUID were removed and replaced with a computed unique ID that is calculated using hash functions and a hash key. While this process is not reversible, it does guarantee that, if another study for the same patient is sent through the pipeline later, those new objects are assigned to the same patient on the research server, unless the hash key in the pipeline is changed. Likewise, additional images/series for the same study would be assigned to the same de-identified study. The MR images and 3D segmentation masks were exported as NIfTI files from the research server using the Python Visage application program interface (API). All images were skull stripped to maintain patient anonymity prior to publication of the dataset.

***Ethical approval.*** The study was conducted according to the guidelines of the Declaration of Helsinki and approved by the Institutional Review Board (or Ethics Committee) of Yale University, protocol 2000029055, approved on 10/01/2020.

## Data Records

The data records are in the process of being published on The Cancer Imaging Archive (TCIA) collections. Each patient has a total of five associated NIfTI files with four image files of the standard sequences (T1 pre-contrast, T1 post-contrast, T2, and FLAIR) and a fifth segmentation file with combined masks from T1 post-contrast and FLAIR segmentations. The segmentation file has three labels: Label 1 (red) represents tumor necrosis, Label 2 (green) represents peritumoral edema, and Label 3 (blue) represents contrast-enhancing tumor. All sequences and segmentations for each patient were exported from research PACS in NIfTI format, co-registered to the SRI24 anatomical template, resampled to a uniform isotropic resolution (1 mm$^3$), and skull stripped. The dataset also contains one Excel file with clinical and qualitative/quantitative imaging feature information for each patient. The patients are labeled with anonymized identifiers.

## Technical Validation

All patients had brain metastases and primary tumor of origin confirmed either pathologically or clinically through the electronic medical record. In addition, only patients with high-quality T1, T1 post-contrast, T2, and FLAIR images without significant motion artifacts were included in the final dataset. All segmentations were independently validated by two neuroradiologists (M.S.A. and F.M.) with more than seven years of clinical experience each. After exporting to NIfTI format, standard sequences and segmentation files for all patients were opened on the ITK-SNAP software and adjusted by a neuroradiologist (M.S.A.).

## Usage Notes

After completion of the data upload process, the NIfTI files can be downloaded from TCIA public collections (https://www.cancerimagingarchive.net/) and opened on segmentation platforms that support NIfTI format.

## Acknowledgements

The authors would like to thank Yale School of Medicine Department of Radiology and Biomedical Imaging and Yale New Haven Hospital for providing the images and helping to make the data publicly available. Research reported in this publication was partly supported by the National Institutes of Health (NIH) under the award number NIH/NCI:U01CA242871 (S.B.). The content of this publication is solely the responsibility of the authors and does not represent the official views of the NIH.

## Author Contributions

D.R. was responsible for data export, publishing the dataset to TCIA, and manuscript preparation. L.J. was responsible for assembling the database, performing tumor segmentations, and collecting clinical and qualitative imaging features for all patients. S.C. was responsible for preparing the final code for lesion counting. A.J., H.M., M.K., G.C.P., S.M., M.v.R. helped with tumor segmentations. F.M. and M.S.A. corrected and validated segmentations. K.B., M.L.,

W.H., and M.W. contributed to image transfer, de-identification, preprocessing, and formatting. N.M., M.S., S.A., S.B., U.B., C.K., and M.D. helped with manuscript revision.

## Competing Interests

M.S.A. has collaborations with Visage Imaging, Inc., Blue Earth Diagnostics, Telix, and AAA. She also has a KL2 TR00186 grant from the NCATS foundation. M.L. is an employee and stockholder of Visage Imaging, Inc., and unrelated to this work, receives funding from NIH/NCI R01 CA206180 and NIH/NCI R01 CA275188. W.H. and M.W. are employees and stockholders of Visage Imaging GmbH. K.B. is an employee of Visage Imaging GmbH. C.K. receives royalties from Primal Pictures 3D Informa, has grant funding from the NIH, and has received the Core Curriculum grant from the American Society of Head and Neck Radiology, all unrelated to this work. The remaining co-authors do not have any competing interests.